# Lone-Pair–Induced Lattice Softness Enables Ultralow thermal conductivity in Hybrid Organic–Inorganic Perovskite GuaPbI3

Rudra P. Singh[1], Shantanu Pathak[2], Saswata Bhattacharya[2], R. Lakshmi Narayan[1*]

[1]Department of Materials Science and Engineering, Indian Institute of Technology Delhi, Hauz Khas, New Delhi, India, 110016

[2]Department of Physics, Indian Institute of Technology Delhi, Hauz Khas, New Delhi, India, 110016

*Corresponding Author: rlnarayan@mse.iitd.ac.in

## ABSTRACT

Thermal conductivity (κ), minimization of inorganic thermoelectrics can only be achieved to a certain extent via nanostructural engineering. Here, we introduce a lone-pair-driven materials design strategy based on chemically induced lattice softness to develop hybrid organic–inorganic perovskites with ultra-low κ. A physics-guided symbolic-regression based machine-learning framework identifies a lone-pair-dominated compositional regime statistically associated with suppressed lattice κ and selects $GuaPbI_3$ as a candidate material. Mechanochemical synthesis yields $GuaPbI_3$ with a room-temperature $\kappa \approx 0.088$ W $m^{-1}$ $K^{-1}$. Electrical measurements reveal electronically active, bias-dependent bulk conduction pathways despite strong phonon suppression, while impedance spectroscopy confirms predominantly bulk-dominated transport behavior. Density functional theory calculations indicate weakly dispersive valence bands, pronounced valence–conduction asymmetry, and localized electrostatic microenvironments arising from charge redistribution within the lattice. Calculated transport coefficients further suggest strong sensitivity of carrier transport to chemical potential, while Lorenz-number analysis reveals deviations from conventional Wiedemann–Franz behavior near band edges. These observations indicate that lone-pair-rich hybrid frameworks generate intrinsically soft and electronically heterogeneous lattice environments capable of strongly suppressing phonon transport while preserving electronically accessible states. This work establishes chemically induced lattice softness as a viable design principle for identifying ultralow-thermal-conductivity hybrid materials without relying on nanostructuring or extrinsic disorder engineering.

## 1. Introduction

Here we introduce a materials-discovery strategy based on chemically induced lattice softness and implement it using a physics-guided symbolic-regression machine-learning framework to identify candidate hybrid perovskites. Rather than functioning as a quantitatively predictive model, the machine-learning approach acts as a compositional screening and ranking framework optimized for sparse literature datasets, enabling physically meaningful trends to emerge across chemically heterogeneous materials spaces. The learned descriptors emphasize lone-pair activity and nonlinear compositional coupling, pointing toward a chemically defined regime of enhanced lattice softness. Application of this framework identifies $GuaPbI_3$ as a representative candidate material. Through synthesis, thermal and electrical characterization, impedance spectroscopy, and first-principles analysis, we show that this material exhibits ultralow lattice thermal conductivity within an ordered crystalline framework together with electronically active bulk transport behavior. More broadly, the results establish chemically programmable lattice softness as a viable route for identifying ultralow-thermal-conductivity hybrid materials without relying on nanostructuring or extrinsic disorder engineering.

Rather than engineering phonon scattering through nanostructuring, this work asks whether intrinsic lattice softness can be identified directly from chemical composition prior to synthesis. Using a physics-guided symbolic-regression framework, we identify a lone-pair-rich compositional regime statistically associated with strongly suppressed lattice thermal conductivity in perovskites. Experimental and computational investigation of $GuaPbI_3$ reveals an electronically active hybrid lattice exhibiting ultralow thermal conductivity together with bias-dependent bulk transport behavior. Electronic-structure calculations further indicate weakly dispersive valence states, pronounced valence–conduction asymmetry, and localized electrostatic heterogeneity within the lattice. Taken together, these observations support the existence of an intrinsically soft and electronically heterogeneous transport environment that differs conceptually from conventional phonon-glass electron-crystal systems.

Thermoelectric materials enable direct conversion between heat and electricity and are widely used in solid-state cooling, temperature sensing, waste-heat recovery, and localized thermal management. Thermoelectric cooling, implemented through thermoelectric coolers (TECs), offers several intrinsic advantages, including the absence of moving parts, compact form factors, high reliability, silent operation, and compatibility with direct-current power sources such as photovoltaic cells and automotive electrical systems. These characteristics make TECs attractive for specialized and emerging applications where mechanical refrigeration is impractical. However, broader deployment remains limited by relatively low energy efficiency and the cost of high-performance thermoelectric materials.

For a thermoelectric module operating between fixed hot- and cold-side temperatures, cooling performance is governed by both the cooling capacity and the coefficient of performance (COP) [1,2]. Under standard assumptions of temperature-independent material properties and equal partitioning of Joule heat between hot and cold sides, the cooling capacity can be expressed as

$$Qc = ST_c I - \frac{1}{2} I^2 R - K\Delta T \qquad (1)$$

where $S$ is the Seebeck coefficient, $I$ is the electrical current, $R$ is the electrical resistance, $K$ is the thermal conductance, and $T_c$ is the cold-side temperature. The maximum achievable coefficient of performance for cooling is given by

$$(COP)_{c,} = \frac{T_c}{T_h - T_c} \cdot \frac{\sqrt{1 + ZT_m} - \frac{T_h}{T_c}}{\sqrt{1 + ZT_m} + 1} \qquad (2)$$

Where $ZT_m$ is the thermoelectric figure of merit evaluated at the average temperature $T_m$. Cooling performance therefore depends on the interplay between entropy transport, electrical conductivity, and heat conduction within the thermoelectric material.

The intrinsic material efficiency of thermoelectrics is characterized by the dimensionless figure of merit

$$ZT = \frac{S^2 \sigma T}{\kappa} \qquad (3)$$

where $S$ is the Seebeck coefficient, $\sigma$ is the electrical conductivity, $T$ is temperature, and $\kappa$ is the total thermal conductivity. For cooling applications, the maximum achievable temperature gradient across a thermoelectric device can be approximated as

$$\Delta T_{max} = \frac{1}{2} Z T_c \quad (4)$$

indicating that reductions in thermal conductivity directly enhance the achievable cooling performance of thermoelectric modules.

Historically, the most widely used thermoelectric materials have been inorganic bulk alloys such as $Bi_2Te_3$, PbTe, SiGe, and $CoSb_3$[3]. Among these, $Bi_2Te_3$-based compounds dominate low-temperature applications below approximately 300 K, while related chalcogenides and antimonides operate at near-room and intermediate temperature regimes. Despite decades of optimization, commercial thermoelectric materials typically exhibit values of $ZT \approx 1$, corresponding to roughly 10% of Carnot efficiency in cooling devices. Theoretical analyses suggest that achieving $ZT \approx 4$ would be required to approach efficiencies comparable to conventional refrigeration systems, yet such values remain difficult to realize in practice.

Following the mid-1990s, significant improvements in thermoelectric performance were achieved through nanostructural engineering. The phonon-glass electron-crystal (PGEC) paradigm emerged as a guiding design principle, aiming to suppress phonon transport while preserving electronic mobility. Strategies including alloy disorder, nanoscale inclusions, quantum wells, quantum wires, and superlattices introduce interfaces that scatter phonons more strongly than electrons, thereby reducing lattice thermal conductivity while maintaining electrical transport[4–8]. Although these approaches increased peak $ZT$ values to approximately 2–3 in research systems, progress has since slowed. Further reductions in lattice thermal conductivity within rigid covalent or ionic inorganic lattices become increasingly difficult without simultaneously degrading carrier mobility.

Organic thermoelectric materials were investigated partly for this reason[9]. Van der Waals–bonded polymers possess inherently low lattice thermal conductivity due to weak intermolecular coupling and structural disorder. Organic semiconductors typically exhibit band gaps in the range of 1–3 eV and require chemical doping to generate mobile carriers. Charge transport commonly occurs through polaron-mediated hopping or mixed metallic-like regimes at high doping levels. However, strong energetic disorder, morphology dependence, anisotropic transport, and a persistent trade-off between Seebeck coefficient and electrical conductivity limit their performance. Despite intrinsically low thermal conductivity, organic thermoelectrics have therefore not achieved competitive cooling performance.

Hybrid organic–inorganic perovskites (HIOPs) occupy an intermediate regime between inorganic and organic thermoelectrics[10–14]. In lead halide perovskites, the electronic band edges are determined primarily by the inorganic Pb–X framework, while the organic A-site cation contributes minimally to the band edges but strongly influences lattice dynamics through rotational degrees of freedom and hydrogen-bonding interactions. These materials are known to exhibit unusually soft lattices, low elastic stiffness, and dense low-energy phonon spectra—characteristics that can strongly suppress phonon transport even in crystalline structures. The coexistence of structural softness and electronically functional band structures therefore suggests a distinct materials platform in which heat conduction may be intrinsically suppressed by chemistry rather than by external nanostructuring.

Despite these characteristics, hybrid perovskites remain comparatively underexplored from the perspective of thermal transport and lattice-softness-driven functionality. Their success in photovoltaic applications—characterized by long carrier lifetimes and defect tolerance—does not automatically imply favorable thermoelectric transport behavior[15–18]. In particular, the relationship between structural softness, electronic transport, and phonon suppression in these systems remains insufficiently understood. Developing chemically guided approaches capable of identifying hybrid lattices with intrinsically suppressed thermal conductivity while preserving electronically accessible states therefore represents an important open challenge.

Here we introduce a materials-discovery strategy based on chemically induced lattice softness and implement it using a physics-guided symbolic-regression machine-learning framework to identify candidate hybrid perovskites[19–22]. Rather than functioning as a quantitatively predictive model, the machine-learning approach acts as a compositional screening and ranking framework optimized for sparse literature datasets, enabling physically meaningful trends to emerge across chemically heterogeneous materials spaces. The learned descriptors emphasize lone-pair activity and nonlinear compositional coupling, pointing toward a chemically defined regime of enhanced lattice softness. Application of this framework identifies $GuaPbI_3$ as a representative candidate material. Through synthesis, thermal and electrical characterization, impedance spectroscopy, and first-principles analysis, we show that this material exhibits ultralow lattice thermal conductivity within an ordered crystalline framework together with electronically active bulk transport behavior. More broadly, the results establish chemically programmable lattice softness as a viable route for identifying ultralow-thermal-conductivity hybrid materials without relying on nanostructuring or extrinsic disorder engineering.

## 2.Methodology

### 2.1. Machine Learning Model

This work develops a machine-learning framework aimed at identifying materials with intrinsically low lattice thermal conductivity. The model employs genetic-programming symbolic regression (GP-SR), which searches directly for compact analytical expressions linking physically motivated descriptors to the target property [23–26]. The Python libraries Gplearn and Joblib were used for implementation[27,28]. Unlike conventional approaches that rely on crystallographic descriptors such as bond lengths or coordination environments, the present framework is based exclusively on composition-derived descriptors. This enables compositional screening of candidate materials prior to crystallographic characterization. Room-temperature lattice thermal conductivity data were compiled from the literature and combined with descriptors constructed from chemical composition. Three physically motivated quantities were defined: $QM$, $GM$, and $LZ$. The first descriptor, QM, represents a quadratic mean of atomic mass and atomic radius across the unit cell, capturing the combined effects of mass contrast and atomic size variation on lattice dynamics:

$$QM = \frac{\sqrt{\sum (Z_i * M_i)^2 + \sum (Z_i * R_i)^2}}{\sqrt{\sum Z_i}} \tag{5}$$

where $Z_i$, $M_i$, and $R_i$ denote the atomic count, atomic mass (amu), and atomic radius (pm) of species i, respectively. The second descriptor, GM, encodes chemical asymmetry through the combined differences in electronegativity and oxidation state between cations and anions:

$$GM = \sqrt{(\chi_{anionaverage} - \chi_{cationaverage}) * (O_{xidationnumberofcation} - O_{xidationnumberofanion})} \quad (6)$$

The third descriptor, LZ, quantifies lone-pair loading within the unit cell:

$$LZ = \left|\sum(Z_i * L_i) - \sum Z_i\right| \quad (7)$$

where $L_i$ denotes the number of lone pairs associated with each atom in the unit cell.
A dataset of 150 lattice thermal conductivity values compiled from 65 studies on perovskite systems was used to construct these descriptors[29–93]. Because the available dataset is comparatively small, GP-SR was selected because it enables identification of compact and physically interpretable functional relationships under explicit complexity constraints. A three-stage evolutionary pipeline was implemented:

- Stage 1 optimizes rank ordering using the Spearman Rank Correlation[94–96]
- Stage 2 improves directional alignment using Cosine Similarity[94,97–100]
- Stage 3 refines distributional agreement using Wasserstein's Distance between KDE distributions[101,102]

Expression complexity was controlled using a parsimony coefficient of 0.015. Populations were propagated between stages to ensure continuity of the evolutionary search. A schematic overview of the algorithm is shown in the supplementary.

## 2.2. Synthesis and XRD

Guanidinium lead iodide ($GuaPbI_3$) was synthesized through solvent-free mechanochemical milling[103]. $PbI_2$ (922 mg) and guanidinium hydroiodide (374 mg) were milled in a Retsch PM100 planetary mill [104] using an 80 mL zirconia jar containing thirty 5 mm zirconia balls (ball-to-powder ratio ≈ 11:1). Milling was performed at 300 rpm for 3 h under dry conditions. Phase formation was examined using powder X-ray diffraction (PANalytical Empyrean, Cu Kα radiation, λ = 1.5406 Å). The obtained diffraction pattern matches the reference pattern of $GuaPbI_3$[105], consistent with phase-pure $GuaPbI_3$ formation within XRD detection limits. The resulting powder was drop-cast onto Si substrates without post-annealing in order to minimize phase evolution following mechanochemical synthesis.

### 2.3. Thermal Characterization

Thermal conductivity was measured using a Linseis Transient Hot Bridge (THB) analyzer[106]. The THB method measures transient heat diffusion using a low-power heat pulse and is particularly suited for ultralow-thermal-conductivity materials because it minimizes errors arising from convection, radiation, and contact resistance. Measurements were conducted using an applied power of 5.0 mW, a current of 10 mA and a measurement duration of 70 s. Each measurement included an initial stabilization period of 5 s followed by repeated thermal-signal acquisition. The sample thickness was approximately 1 mm. Temperature drift during measurement was monitored and remained small (average 0.0196 °C $s^{-1}$). Measurements that exceeded instrument temperature limits were automatically flagged and excluded from analysis.

### 2.4. Electrical Characterization

Current–voltage measurements were conducted using a Keithley 8009 high-voltage test fixture[107]. Measurements were performed under applied voltages ranging from 3 V to 2 kV, enabling exploration of both low- and high-power regimes. Impedance spectroscopy was performed using a silver-coated pellet (~1 mm thickness). An AC excitation of 1 V was applied over a frequency range of 0.03 Hz–1 GHz. Measurements were carried out in the temperature range –99 °C to 80 °C, corresponding to conditions relevant for low-temperature transport characterization and material stability.

### 2.5. Density Functional Theory

Density-functional-theory calculations were performed using VASP 5.4.4 to investigate the electronic structure and transport-relevant features of $(Gua)PbI_3$[108–111]. The projector augmented-wave (PAW) method was employed together with the Perdew–Burke–Ernzerhof (PBE) exchange–correlation functional[112]. Although PBE underestimates absolute band gaps, it provides reasonable descriptions of band dispersion, orbital character, and relative energy alignment, which are the focus of this work. Structural optimization was carried out until the total-energy change was below $10^{-5}$ eV and residual atomic forces were less than $10^{-4}$ eV Å$^{-1}$. A plane-wave cutoff energy of 520 eV was used, and Brillouin-zone integration was performed using a 4 × 8 × 2 Monkhorst–Pack k-point mesh. To account for heavy elements, all calculations included fully relativistic spin–orbit coupling and non-collinear magnetization (LSORBIT = TRUE, LNONCOLLINEAR = TRUE). Band structures and densities of states were analyzed in absolute energy coordinates. The valence-band maximum (VBM) and conduction-band minimum (CBM) were determined from the density of states using a finite threshold to suppress numerical noise. Energy windows were then defined to identify valence- and conduction-frontier states. Because strong spin–orbit coupling leads to band mixing, individual bands were treated independently, and their dispersion was quantified using the per-band bandwidth:

$$\Delta E_b = max_k E_b(k) - min_k E_b(k) \quad (8)$$

Bands were classified as valence- or conduction-frontier states if they intersected the corresponding energy windows. In addition to reciprocal-space analysis, real-space electronic fields were evaluated to probe spatially resolved electronic behavior. Electrostatic potential, charge density, and derived quantities were obtained from self-consistent calculations and analyzed on three-dimensional grids using VESTA[113]. Transport coefficients ($\sigma/\tau$, S, $S^2\sigma/\tau$, $\kappa_e$) were calculated using BoltzTraP 1.2.5 within the constant relaxation-time approximation at 300 K and 373 K. The calculated transport coefficients should be interpreted qualitatively within the constant relaxation-time approximation. Further details regarding real-space analysis and transport calculations are provided in the Supplementary Information.

# 3. Results

## 3.1. Machine Learning Model

The symbolic-regression framework converged to a stable analytical expression across multiple independent runs. The resulting ranking function is:

$$F(QM, GM, LZ) = 23.882 + GM + e^{LZ} - (LZ * QM)^{\frac{1}{3}} \qquad (9)$$

Higher values of $F$ correspond to compositions statistically associated with lower lattice thermal conductivity. Rather than performing direct regression, the model functions as a ranking tool that identifies compositional regimes enriched in ultralow-$\kappa_l$ candidates. Application of this expression to an independent dataset [114] of hybrid organic–inorganic perovskites identified $GuaPbI_3$ as a top-ranking material**[Fig. 1(a)]**, which was subsequently selected for experimental validation. The reproducibility of the functional form across independent runs indicates convergence toward a stable compositional trend rather than a run-specific artifact. Additionally, analysis of the lone-pair descriptor reveals a strong inverse relationship between $\log LZ$ and $\log \kappa_l$, with $R^2$=0.58**[Fig. 1(b)]**. A transition is observed near $LZ \approx 6$, beyond which thermal conductivity decreases approximately inversely with $LZ$, suggesting the emergence of a lone-pair-dominated lattice-softness regime. Collectively, these results establish $F$ as a robust composition-based screening tool for identifying materials with low lattice thermal conductivity. Further validation details are provided in the Supplementary Information.

**Fig. 1. Symbolic-regression screening for low lattice thermal conductivity.**

Fig. 1. (a) The ranking function identifies $GuaPbI_3$ as a top candidate among hybrid organic–inorganic perovskites, (b) the inverse log LZ–log $\kappa_l$ trend suggests a lone-pair-dominated lattice-softness regime beyond LZ ≈ 6.

(a)

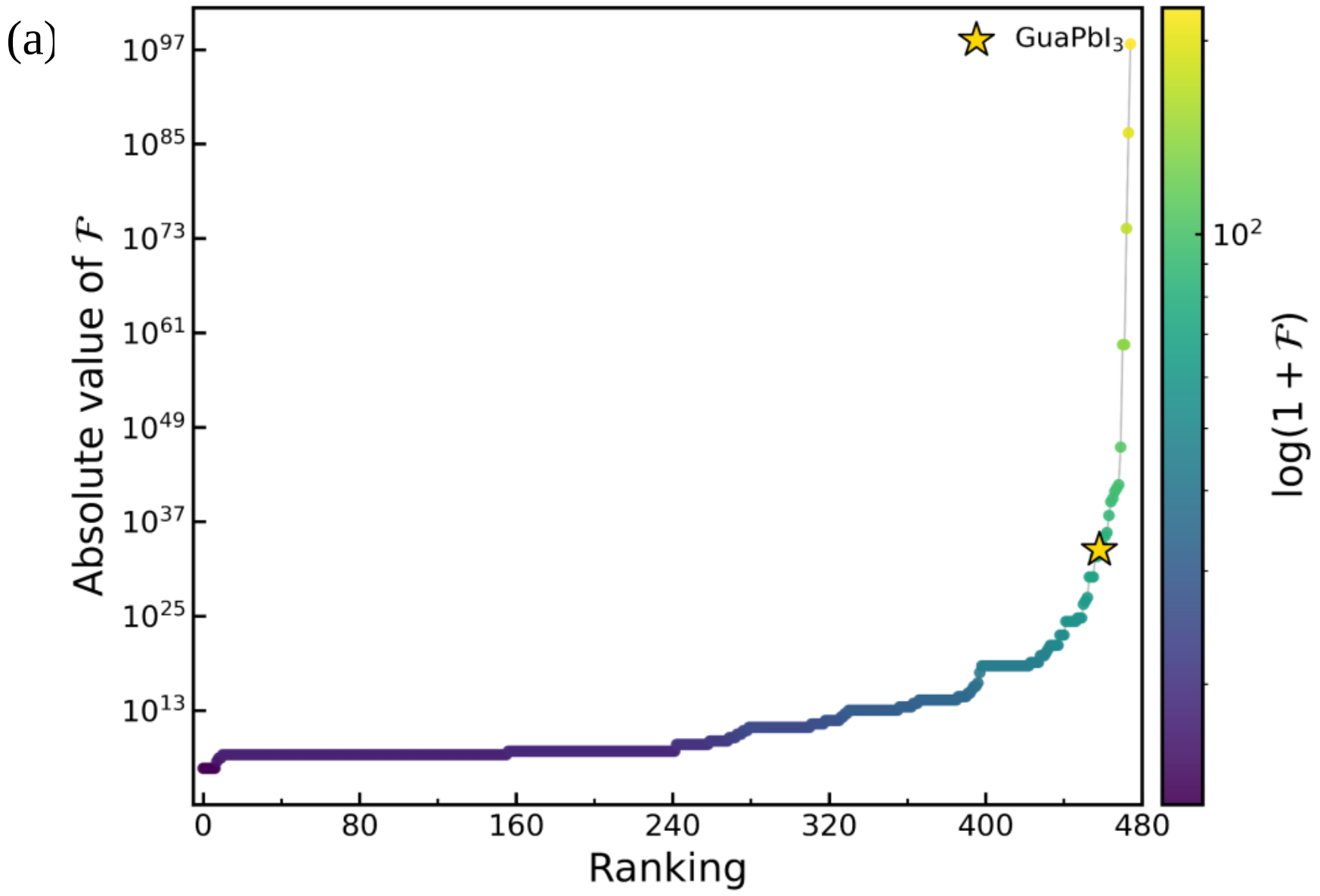

(b)

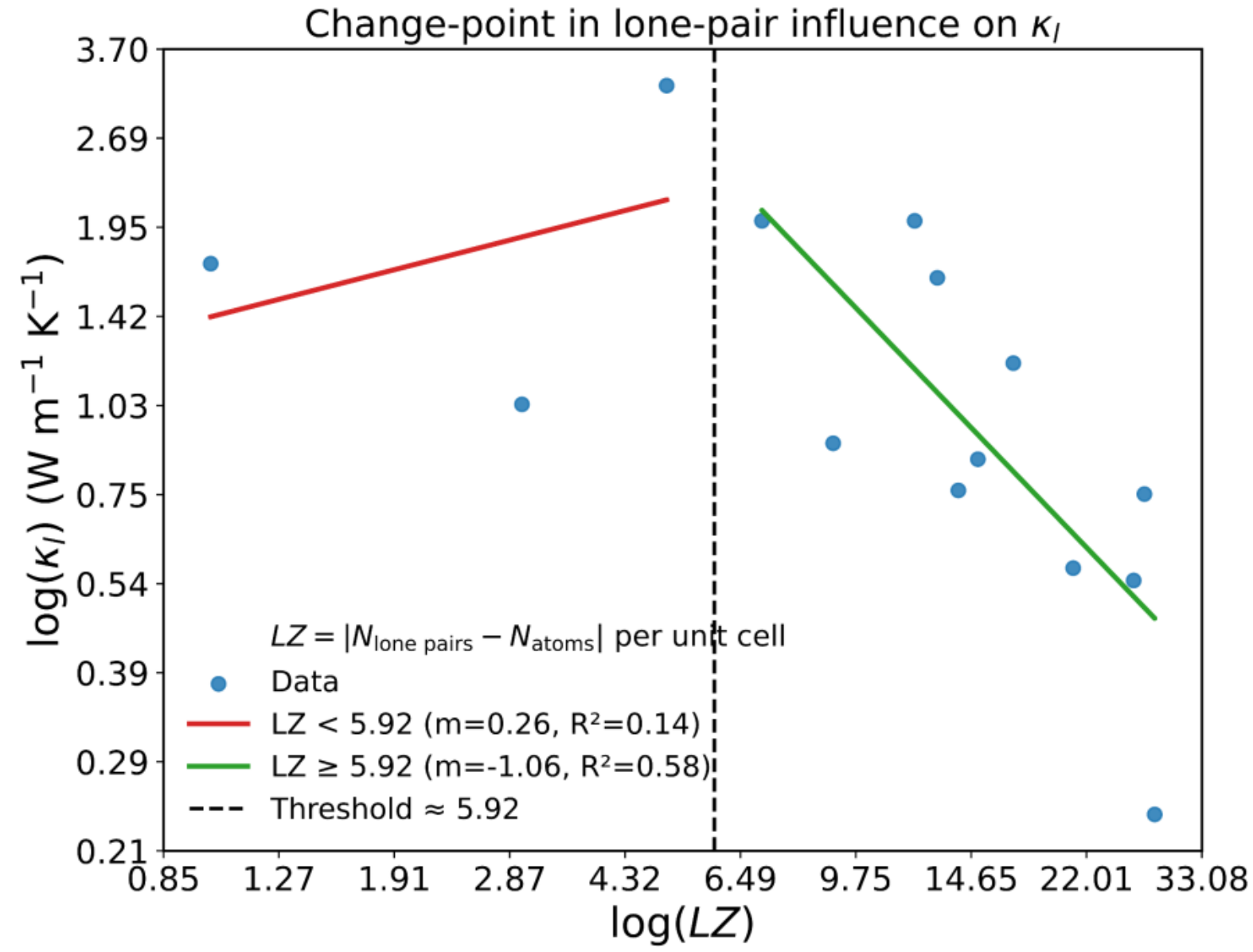

### 3.2. Synthesis and XRD

The XRD pattern confirmed the formation of the GuaPbI3 compound as all the peaks were present. The diffraction peaks at 2θ ≈ 7.5°, 12°, and 15° match the reference pattern of orthorhombic $GuaPbI_3$, confirming successful phase formation without detectable secondary phases.

**Fig. 2. XRD pattern of obtained GuaPbI3 sample**

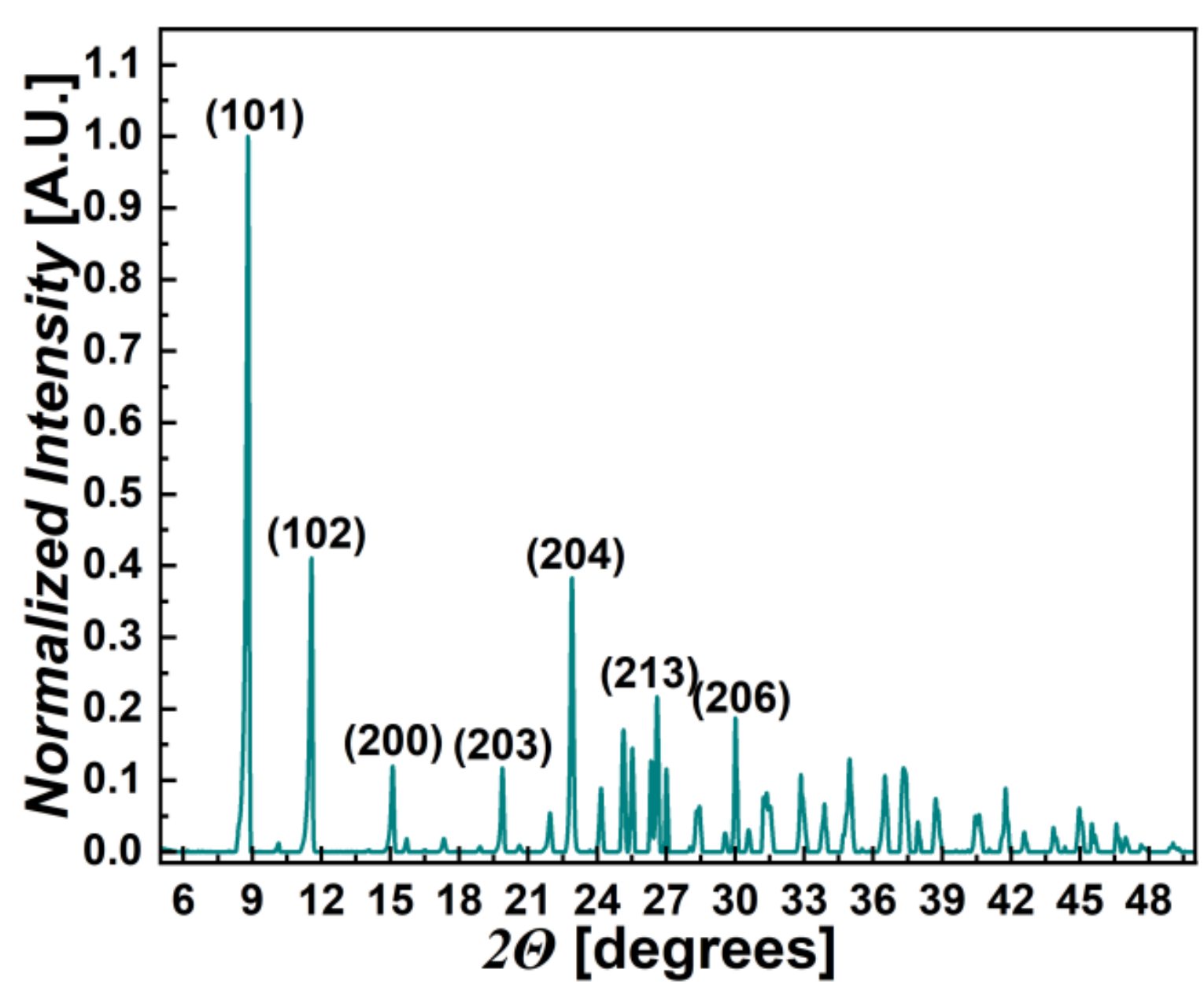

### 3.3. Thermal Characterization

Transient thermal characterization of the $GuaPbI_3$ thin film at 32 °C yields a thermal conductivity of 0.088 W $m^{-1}$ $K^{-1}$. This value is more than an order of magnitude lower than that of benchmark thermoelectric materials such as $Bi_2Te_3$ (≈ 1.2–1.6 W $m^{-1}$ $K^{-1}$), placing $GuaPbI_3$ firmly within the regime of ultralow thermal conductivity materials. To our knowledge, this $\kappa_l$ value is among the lowest reported for crystalline hybrid perovskites at room temperature and places $GuaPbI_3$ within the regime of ultralow-thermal-conductivity crystalline hybrid materials reported at room temperature.

### 3.4. Electrical Characterization

### *3.4.1. I-V Characterization*

Electrical measurements were performed under fixed applied voltages of 2000 V and 3000 V, corresponding to distinct applied-bias conditions defined by the electrical power P=IV. Conductance was calculated from the measured volume resistivity using an effective film thickness of ~1 mm and an active area of 1 cm². The relationship between current and conductance shows a linear dependence over the measured range, with the data following a straight line passing through the origin. The fitted slope is approximately unity, corresponding to a near Y=X relationship between current and conductance **[Fig. 3(a)]**. The conductance also varies linearly with applied power**[Fig. 3 (b)]**, within the measured range. The conductance–power relationship follows the same proportional trend observed in the current–conductance data. The primary results are given in supplementary.

**Fig. 3. Electrical transport scaling under high bias**

Fig. 3. (a) Current and (b) applied power show near-linear proportionality with conductance across 2000–3000 V. The Y ≈ X trend indicates Ohmic-like conductance scaling within the measured range.

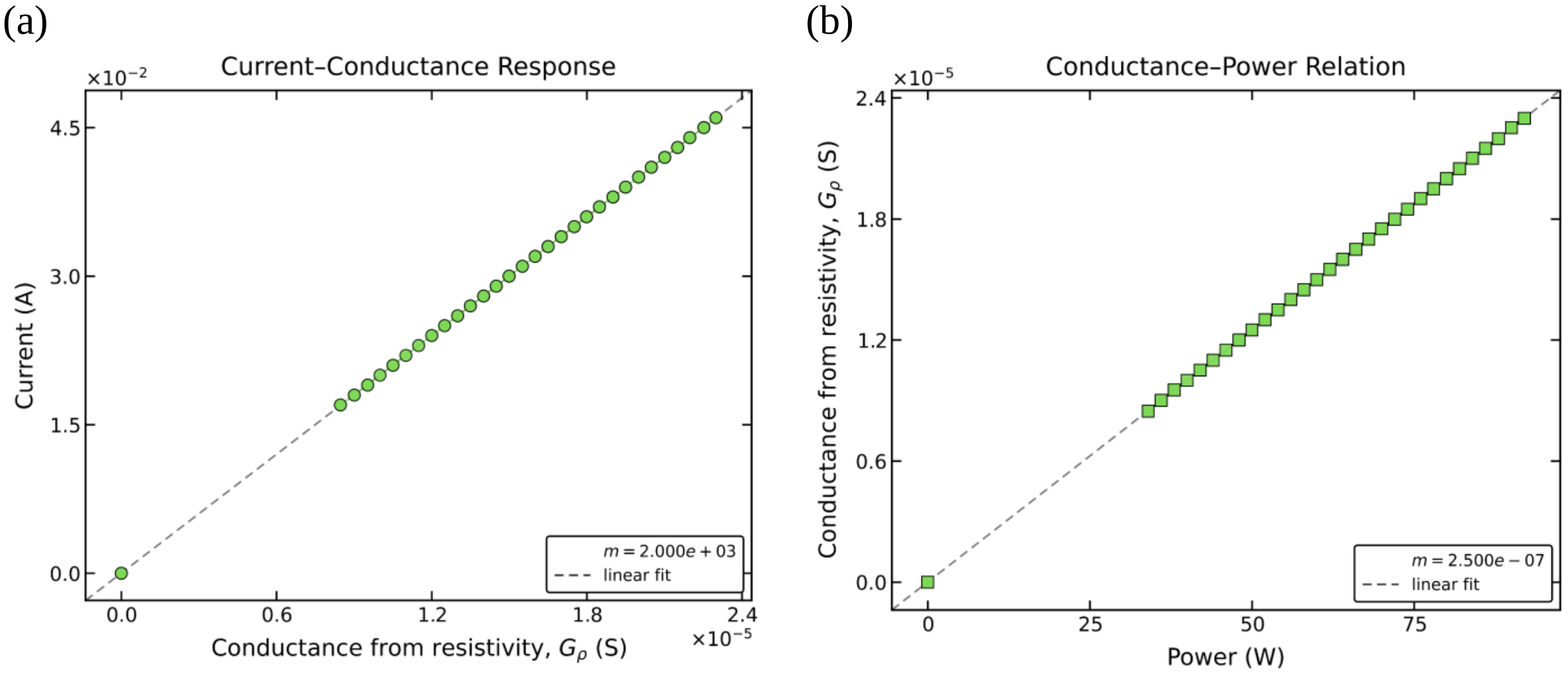

### *3.4.2. Impedance Spectroscopy*

The loss tangent (tan $\delta$) and phase tangent (tan $\Phi$) **[Fig. 4 (a), (b)]** exhibit systematic dependence on frequency and temperature. Both quantities show strong variation in the low-frequency regime, particularly near ~0.05 Hz, where peak magnitudes are observed. At higher frequencies, both tan $\delta$ and tan $\Phi$ decrease in magnitude and display reduced sensitivity to temperature. The magnitude of both quantities increases with temperature in the low-frequency regime, with the highest values observed near 79.6 °C. Polarization **[Fig. 4 (c)]** follows a similar frequency-dependent trend, with higher values at low frequencies and a progressive decrease with increasing frequency. The temperature dependence of polarization remains consistent across the measured range, with higher temperatures corresponding to increased polarization magnitude at low frequencies. Results on primary physical quantities are discussed in the supplementary.

**Fig. 4. Temperature–frequency dependent impedance response of GuaPbI3**

Fig. 4. (a) tan $\delta$, (b) tan $\Phi$, and (c) polarization are strongest at low frequency, especially near ~0.05 Hz, and increase with temperature. Their suppression at higher frequency indicates slow, thermally activated dielectric relaxation involving charge/dipolar polarization.

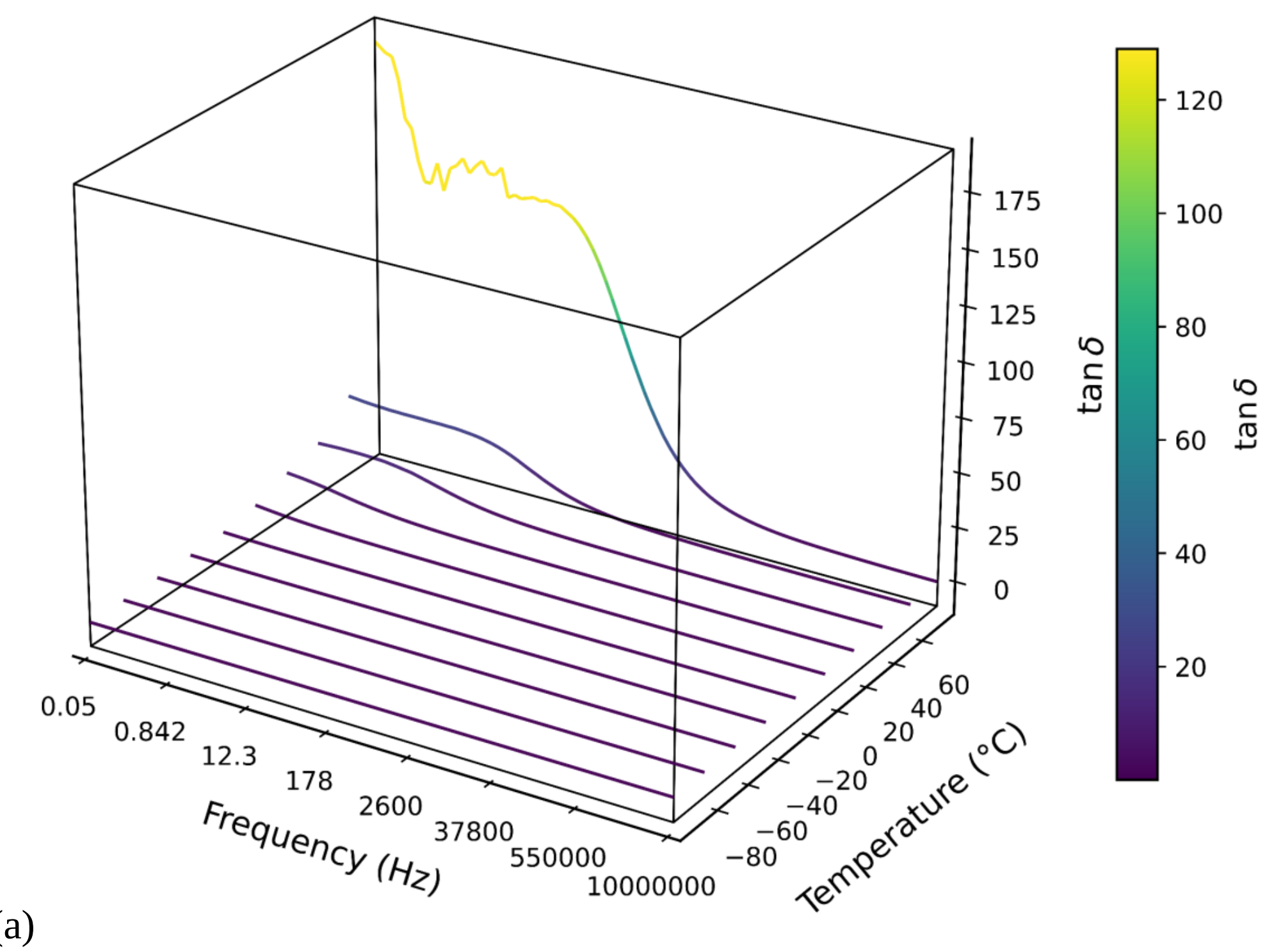

(a)

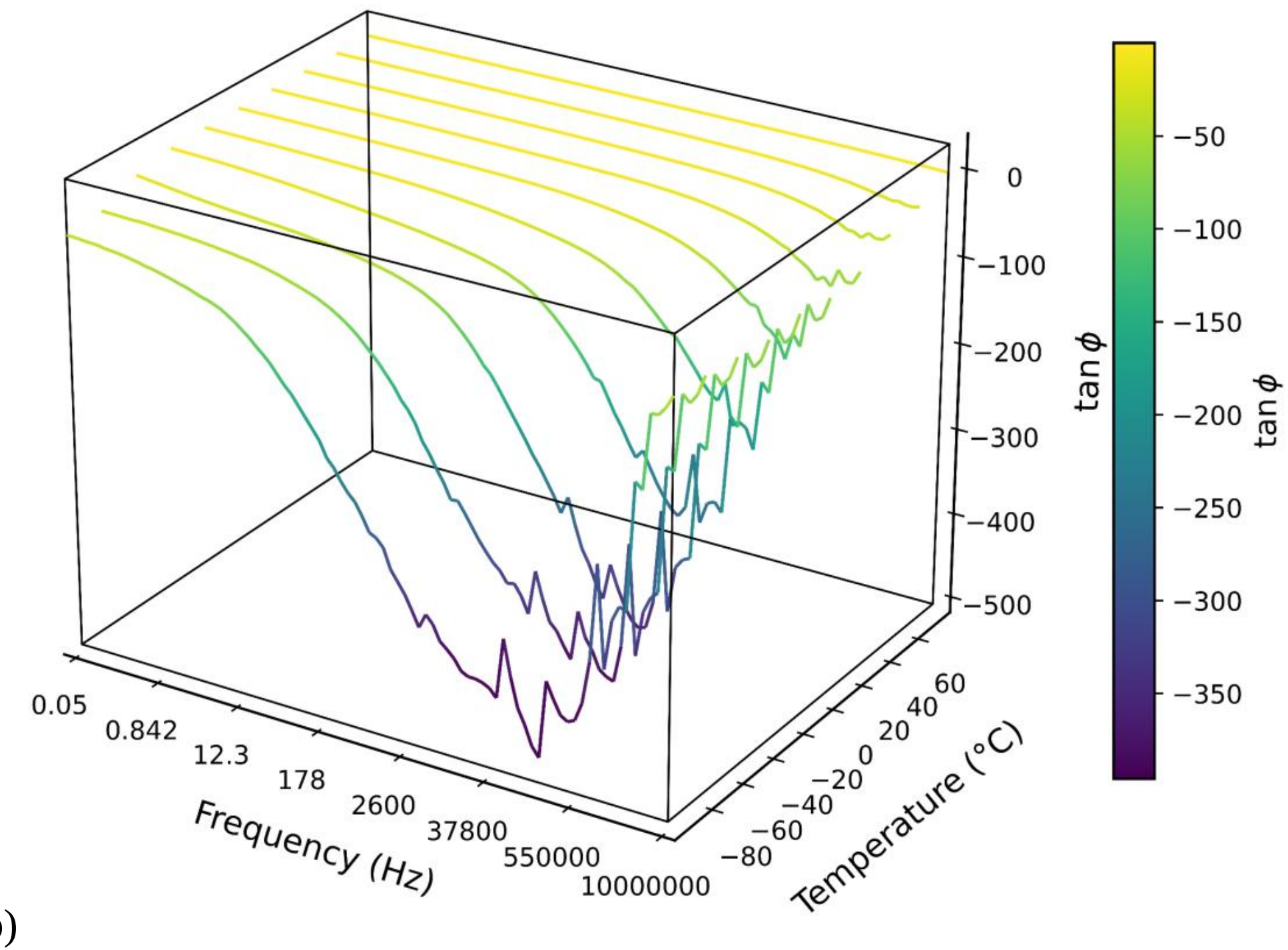
tan $\phi$
tan $\phi$
Frequency (Hz)
Temperature (°C)
0.05
0.842
12.3
178
2600
37800
550000
10000000
−80
−60
−40
−20
0
20
40
60
0
−100
−200
−300
−400
−500
−50
−100
−150
−200
−250
−300
−350

(b)

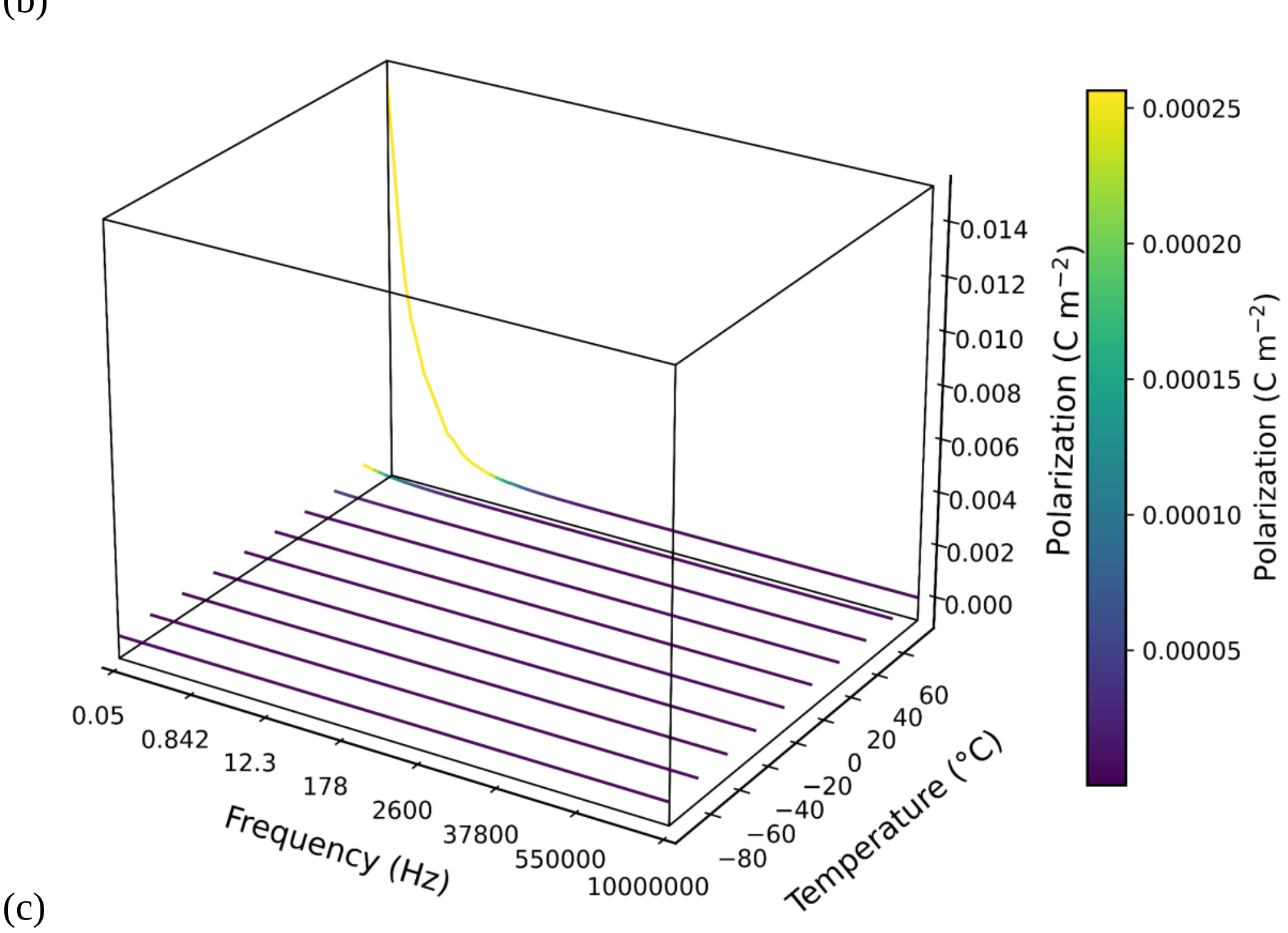
Polarization (C m$^{-2}$)
Polarization (C m$^{-2}$)
Frequency (Hz)
Temperature (°C)
0.05
0.842
12.3
178
2600
37800
550000
10000000
−80
−60
−40
−20
0
20
40
60
0.014
0.012
0.010
0.008
0.006
0.004
0.002
0.000
0.00025
0.00020
0.00015
0.00010
0.00005

(c)

### 3.5. Density Functional Theory

Frontier-band analysis **[Fig. 5 (a)]** reveals a strong asymmetry between valence- and conduction-band dispersions. The valence frontier consists of 24 bands with very low dispersion, with a median bandwidth of $\Delta E \approx 0.007$ eV and a maximum of $\Delta E \approx 0.015$ eV, with all bands satisfying $\Delta E \leq 0.02$ eV. In contrast, the conduction frontier contains 16 bands with significantly higher dispersion, with a median $\Delta E \approx 0.19$ eV and a maximum $\Delta E \approx 0.49$ eV, and only ~25% of bands satisfying $\Delta E \leq 0.10$ eV. Valence-band dispersion is primarily confined near the Γ-point, coinciding with a high density of states near the band edge. Real-space potential mapping **[Fig. 5 (b)]** shows strong internal modulation across the unit cell, with multiple local extrema distributed throughout the lattice and within microcavity regions. Deformation-density maps **[Fig. 5 (c)]** indicate lattice-wide charge redistribution, with enhanced response in microcavities. The electron-density gradient magnitude **[Fig. 5 (d)]** resolves a sparse but connected network of sharp internal electronic interfaces extending across the unit cell. These high-gradient regions occupy only a small fraction of the total volume but are distributed throughout the structure. The Seebeck coefficient **[Fig. 6 (a), (b)]** shows strong dependence on chemical potential and temperature. The calculated Seebeck profiles shift toward $\mu \approx 0$ eV with increasing carrier concentration, approaching within ~0.5 eV of the Fermi level. Additional satellite features emerge near $\mu \approx \pm 0.1$ eV. With increasing temperature, the Seebeck profile becomes more sharply centered around $\mu \approx 0$ eV, with enhanced peak definition. The calculated Lorenz number $L = \kappa_e/(\sigma T)$ **[Fig. 6 (c), (d)]** exhibits a pronounced central region near $\mu \approx 0$ eV where the ratio becomes ill-defined due to vanishing $\sigma$ and $\kappa_e$. At 300 K, this region extends approximately from −0.069 to +0.070 eV, and narrows slightly at 373 K to about −0.069 to +0.063 eV. Outside this region, L remains below the Sommerfeld value across a wide chemical-potential range before approaching it at higher carrier concentrations. In the doped system, L remains finite across the central region and shows multiple excursions above the Sommerfeld value, including spike-like features at larger chemical potentials. These calculated transport quantities should be interpreted qualitatively within the constant relaxation-time approximation. Full band structure, density of states, and additional transport quantities are provided in the Supplementary Information.

**Fig. 5. Frontier electronic structure and internal charge landscape**

Fig. 5. (a) Flat valence-frontier bands contrast with more dispersive conduction bands, indicating asymmetric carrier localization, (b) potential, (c) deformation-density, and (d) density-gradient maps reveal microcavity-enhanced charge redistribution and sparse connected electronic interfaces.

(a)

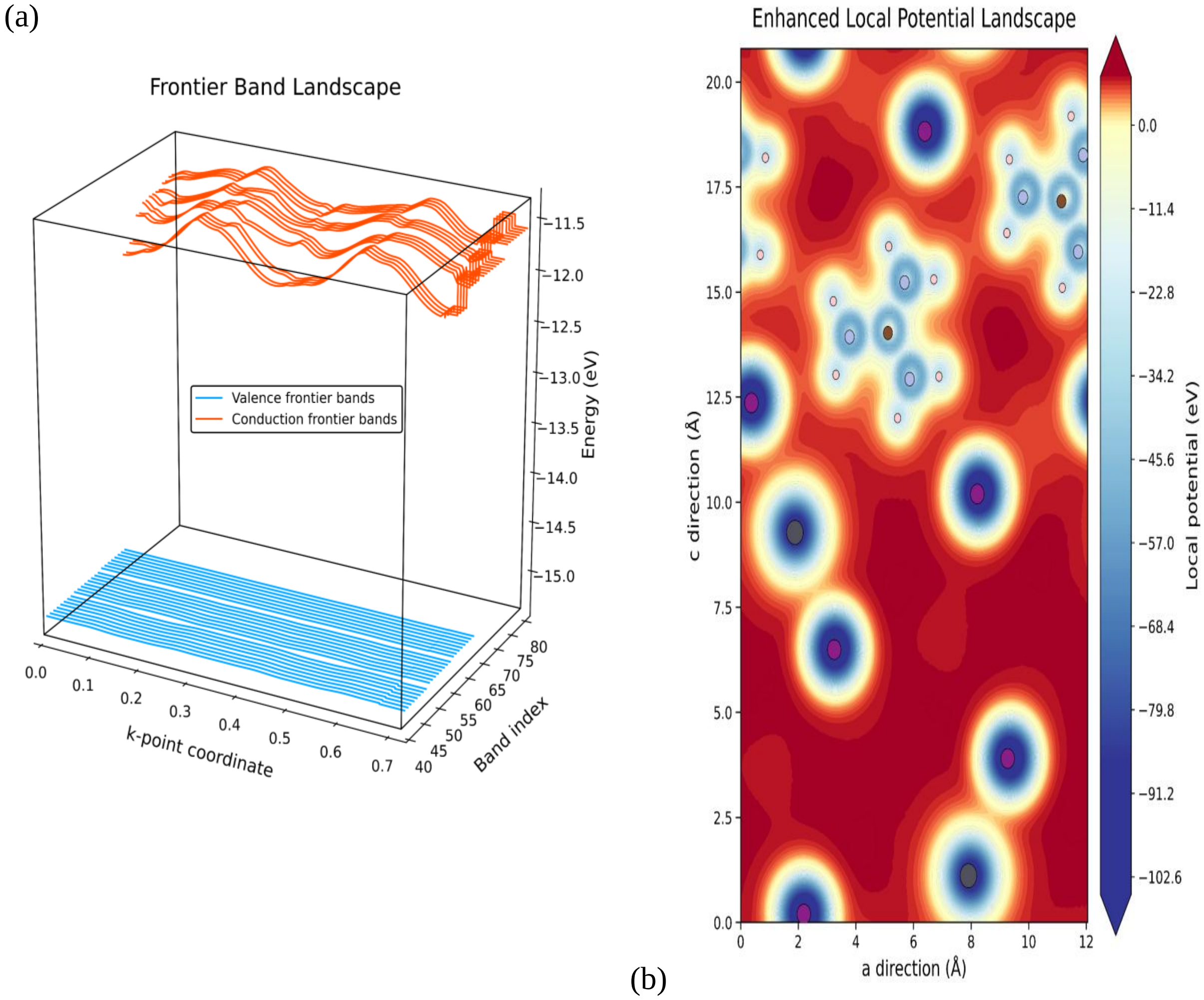

(b)

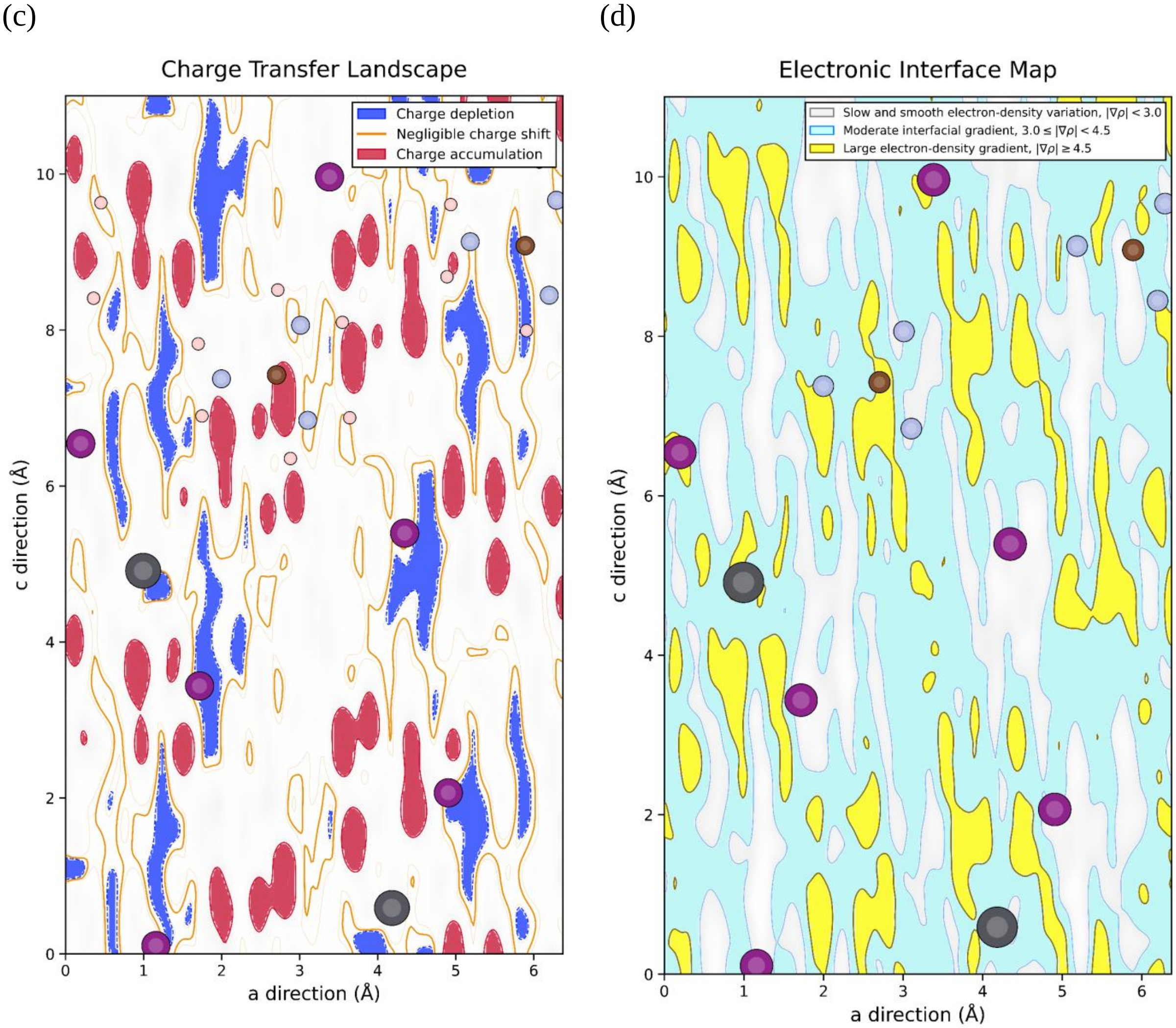
(c)
Charge Transfer Landscape
Charge depletion
Negligible charge shift
Charge accumulation
c direction (Å)
a direction (Å)
(d)
Electronic Interface Map
Slow and smooth electron-density variation, |∇ρ| < 3.0
Moderate interfacial gradient, 3.0 ≤ |∇ρ| < 4.5
Large electron-density gradient, |∇ρ| ≥ 4.5
c direction (Å)
a direction (Å)

**Fig. 6. Calculated thermoelectric transport response at 300K and 373K respectively**

Fig. 6. (a)-(b) The Seebeck coefficient shifts toward the Fermi level with carrier concentration and becomes more sharply defined with temperature, (c)-(d) the Lorenz number remains non-Sommerfeld-like near low-conductivity regions, indicating qualitative deviations from simple metallic transport.

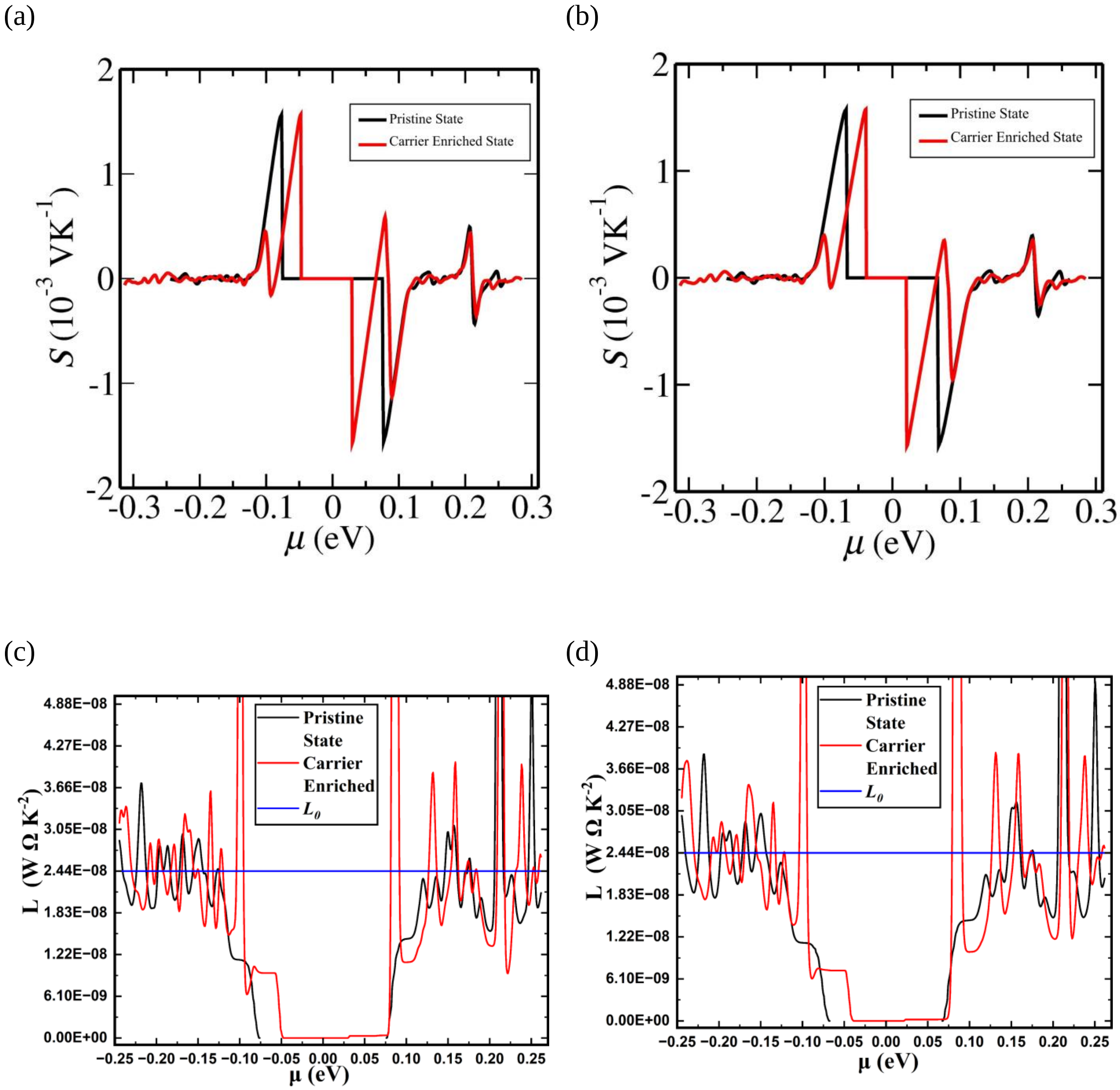

**Fig. 7. Ashby Chart**

Comparative transport-property positioning of $GuaPbI_3$ relative to representative thermoelectric material classes and $MAPbI_3$, which is from the same HIOP family[115–122]

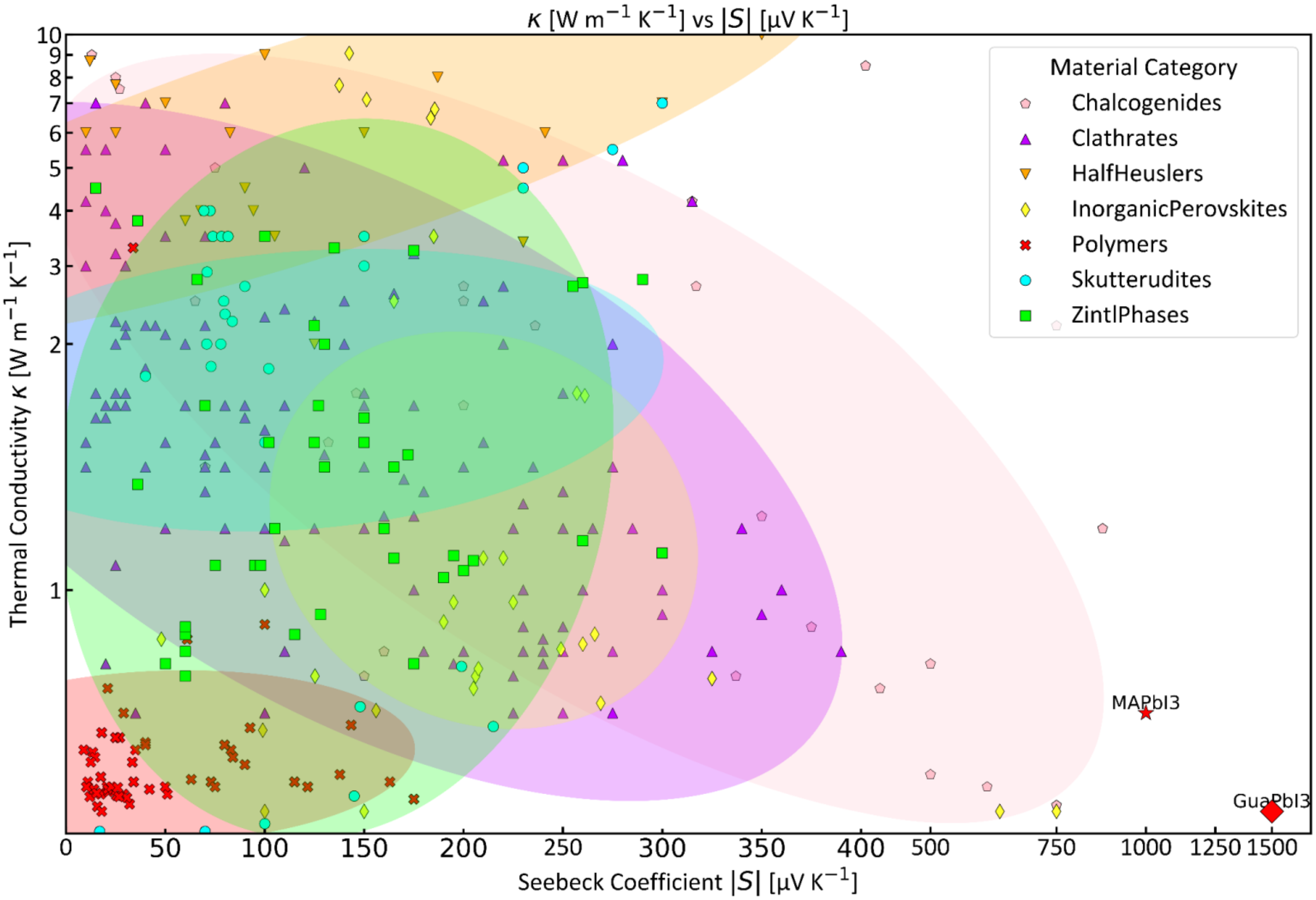

Taken together, the experimental and computational results indicate that $GuaPbI_3$ combines ultralow lattice thermal conductivity with electronically active bulk transport behavior within a structurally soft hybrid framework.

## 4. Discussion

Rather than investigating whether thermoelectric performance can be enhanced through engineering approaches such as nanostructuring or interface scattering to reduce lattice thermal conductivity, the present study addresses a different question: whether materials possessing intrinsically low lattice thermal conductivity ($\kappa_l$) can be identified prior to synthesis. The underlying premise is that chemical composition itself can reveal intrinsic lattice softness capable of strongly suppressing phonon transport. A physics-guided machine-learning framework based upon genetic programming and symbolic regression was developed to explore this possibility by constructing a ranking operator $F$ which correlates with lattice thermal conductivity. The operator functions as a ranking tool rather than as an absolute predictor and utilizes only the chemical composition in order to predict the relative ordering of compounds according to decreasing $\kappa_l$. The training dataset included structurally heterogeneous perovskites encompassing oxides, halides, chalcogenides, inverse perovskites, double perovskites, and related structures, and the evolutionary search converged to the same symbolic functional motif across multiple randomized runs. This indicates that the resulting expression represents a stable compositional trend rather than a run-specific artifact. A regime transition appears near $LZ \approx 6$, beyond which lattice thermal conductivity at 300 K decreases approximately inversely with the lone-pair descriptor $LZ(R^2 \approx 0.58)$. Its persistence across structurally diverse perovskites suggests that lone-pair density $\left(LZ = N_{lonepair\ e_s^-} - N_{atoms}\, perunitcell\right)$ acts as a strong indicator of lattice softness within this compositional space despite the relation being statistical rather than universal.

The next question is therefore how this lone-pair-rich compositional signature manifests physically in $GuaPbI_3$ itself. The combined experimental and computational observations suggest the presence of a highly heterogeneous lattice–charge environment generated by the lone-pair-rich Pb–I framework together with the large polarizable guanidinium cation. Within such an environment, local lattice distortions are expected to respond strongly to electrostatic perturbations and may interact significantly with electronically active states. This picture provides a possible explanation for the coexistence of ultralow lattice thermal conductivity and electronically active bulk transport behavior observed experimentally. The following discussion therefore examines whether the structural, electronic, and transport observations are collectively consistent with a strongly coupled and dynamically deformable transport environment.

The first step is therefore to examine the structural origin of the strongly suppressed phonon transport[123–127]. The coexistence of ultralow lattice thermal conductivity with electronically active transport provides the first indication of this unusual transport environment. The measured thermal conductivity $\kappa_{total} \approx 0.088\ Wm^{-1}K^{-1}$ lies more than an order of magnitude below that of conventional thermoelectric materials such as $Bi_2Te_3$. However, electrical measurements indicate that bulk conduction pathways remain active under applied bias despite this strong phonon suppression. In crystalline materials strong phonon scattering often accompanies structural disorder which simultaneously suppresses electronic conduction, making such coexistence unusual. In the present system, however, phonon transport appears to be suppressed much more strongly than charge transport while the lattice remains crystalline.

From kinetic theory,

$$\kappa_l = \frac{1}{3} C_{ph} v_{avg}^2 \tau, \quad (10)$$

where $C_{ph}$ is the phonon heat capacity, $v_{avg}$ is the average phonon group velocity, and $\tau$ is the phonon lifetime. Ultralow lattice thermal conductivity implies strongly reduced phonon lifetimes or mean free paths. Hybrid halide perovskites are already known to exhibit relatively low sound velocities and strong lattice anharmonicity compared with conventional inorganic semiconductors. Low phonon group velocities together with strong lattice anharmonicity reduce the energetic cost of local lattice distortions, thereby facilitating dynamic carrier–lattice coupling and potentially favoring localized or large-polaron-like carrier states.

The introduction of a large organic cation further increases structural flexibility within the lattice. Guanidinium interacts with the Pb–I framework primarily through hydrogen bonding and Coulomb interactions rather than through direct orbital hybridization with the band edges and occupies a substantial fraction of the perovskite A-site cavity. Hence, the organic cation acts as a structural perturbation embedded within the inorganic framework. Guanidinium is also highly polarizable and possesses multiple internal vibrational modes, resulting in a dynamic response to local electrostatic fields generated by mobile charges[128]. The ion exhibits a relatively large effective radius (r ≈ 0.21 nm) and high electronic polarizability ($\alpha \approx 4.44 \times 10^{-30}$ m$^3$), indicating that the organic cation can be readily distorted by local electrostatic fields within the lattice. Its thermodynamic properties further reflect a high density of accessible internal degrees of freedom, with a molar entropy of approximately $S \approx 264$ J K$^{-1}$ mol$^{-1}$ and a constant-pressure heat capacity $C_p \approx 78$ J K$^{-1}$ mol$^{-1}$ at 298 K. These properties imply numerous accessible rotational and vibrational configurations of the guanidinium ion. Consequently, the Pb–I lattice can undergo persistent local distortions through simultaneous electrostatic and hydrogen-bond interactions with the guanidinium ion. The electronic charge distribution of guanidinium can therefore respond dynamically to local electrostatic fields, enabling the organic sublattice to generate fluctuating electrostatic microenvironments throughout the crystal.

These dynamic electrostatic fluctuations may provide an environment favorable for enhanced carrier–lattice interactions, particularly in regions where local electrostatic heterogeneity is significant.

Charge-density-difference analysis further reveals localized regions of charge accumulation and depletion distributed throughout the lattice. These regions form electrostatic microenvironments that appear as microcavity-like pockets in the charge-transfer landscape. Although their precise structural origin remains uncertain, they represent internal electronic interfaces across which the electrostatic potential varies sharply. Such electrostatic heterogeneity introduces spatially varying bonding environments that can act as effective phonon-scattering centers while preserving overall crystallinity. The presence of these charge-transfer microenvironments also provides a physical interpretation for the lone-pair descriptor $LZ$ identified by the machine-learning model. Lone-pair-active ions generate highly polarizable and spatially anisotropic electron densities that naturally produce localized regions of charge accumulation and depletion. In $GuaPbI_3$, the combination of the lone-pair-active $Pb^{2+}$ framework and the large polarizable guanidinium cation therefore creates a strongly heterogeneous electrostatic landscape capable of interacting strongly with both phonons and charge carriers.

Electronic structure calculations provide additional insight into this transport environment. The valence band manifold exhibits extremely weak dispersion, with median bandwidths of approximately $\Delta E \approx 0.007\ eV$, indicating high electronic polarizability and strong sensitivity of the electronic states to local lattice distortions. In contrast, pronounced asymmetry between the valence and conduction states is observed because the conduction band manifold is comparatively more dispersive. Substantial electrostatic heterogeneity within the unit cell is observed through real-space electronic analysis. Electrostatic potential maps and deformation density plots reveal spatial modulation of the electronic environment across the lattice, while electron-density gradient analysis identifies localized regions of strong gradients associated with microcavity-like electronic features. Together, these structural and electronic features indicate a strongly heterogeneous electrostatic landscape that is likely to influence carrier motion and localization behavior.

Electrical measurements are consistent with this complex transport environment. A transition in the electrical response regime indicates that stable conduction pathways emerge once sufficient bias activates preferred transport channels. At low applied bias, current and resistivity exhibit irregular temporal fluctuations suggesting unstable conduction pathways. At higher bias, however, current increases while resistivity decreases, producing an approximately linear current–conductance relationship. Impedance spectroscopy further indicates that conduction remains predominantly bulk dominated rather than surface limited.

Although direct thermoelectric measurements such as Seebeck coefficient and ZT were not experimentally obtained in the present study, the combination of ultralow thermal conductivity and electronically active transport behavior motivates consideration of the possible thermoelectric implications of this transport environment. Calculated transport coefficients indicate substantial sensitivity of the Seebeck response to carrier concentration and chemical potential, suggesting that moderate electronic conductivity together with strong phonon suppression could be relevant for future thermoelectric exploration.

These transport features also influence bipolar conduction. Because electrons and holes possess opposite Seebeck signs, their contributions partially cancel according to

$$S = \frac{\sigma_n S_n + \sigma_p S_p}{\sigma_n + \sigma_p} \tag{11}$$

while bipolar heat transport contributes an additional thermal conductivity term

$$\kappa_{bi} = \frac{\sigma_n \sigma_p}{\sigma_n + \sigma_p} \left(S_n - S_p\right)^2 T \tag{12}$$

In the present system these effects remain limited because the relatively large band gap (~2.3 eV) suppresses thermal excitation of minority carriers, while the asymmetry between valence and conduction band dispersions may further reduce minority carrier mobility.

The relationship between electrical conductivity and electronic thermal conductivity must also be considered. In conventional semiconducting materials these quantities are linked through the Wiedemann–Franz law

$$\kappa_e = L\sigma T \tag{13}$$

First-principles transport calculations provide access to the Lorenz number as a function of chemical potential. In the undoped system the Lorenz-number curve exhibits a central gap around $\mu \approx 0$, corresponding to a region where both electrical conductivity and electronic thermal conductivity become extremely small, rendering the ratio $L = (\kappa_e/\sigma T)$ ill-defined. At 300 K this region extends approximately from −0.069 to +0.070 eV and narrows slightly at 373 K. Outside this intrinsic window the Lorenz number remains below the Sommerfeld value across a substantial chemical-potential range before approaching $L_0$ at larger carrier concentrations. Such suppression of the Lorenz number relative to the Sommerfeld value indicates that electronic heat transport does not scale proportionally with electrical conductivity and reflects the strong energy dependence of carrier transport near the band edges.

In the doped system the Lorenz number remains finite throughout the central chemical-potential region but exhibits frequent excursions above the Sommerfeld value, including spike-like enhancements. These features likely arise from the pronounced asymmetry between the valence and conduction band dispersions together with increasing contributions from minority carriers at elevated chemical potentials. The persistence of Lorenz values below the Sommerfeld value over a broad chemical-potential range is also consistent with energy-selective transport near the band edges, in which carriers contributing most strongly to electrical conduction do not contribute proportionally to electronic heat transport. Such behavior further indicates that the relationship between heat and charge transport deviates from the conventional Wiedemann–Franz picture.

The structural characteristics of $GuaPbI_3$ may also be relevant for future doping studies. The large unit-cell volume creates spacious structural cavities within the lattice. Coulomb interactions between charged defects scale approximately as

$$V(r) = \frac{q_1 q_2}{4\pi\varepsilon r}, \tag{14}$$

Thus, increasing the effective separation between charges reduces electrostatic binding energies. Together with dielectric screening provided by the polarizable guanidinium cation, this environment may weaken Coulomb trapping of dopant ions and facilitate carrier introduction without strong localization.

Taken together, these observations are consistent with a transport environment in which carrier motion is likely influenced by dynamically deformable lattice distortions within a heterogeneous electrostatic landscape. In this picture, local lattice distortions and electrostatic microenvironments may interact strongly with electronically active states, contributing to the strong suppression of coherent phonon transport while allowing bias-activated bulk conduction pathways to persist. The resulting behavior may share features with large-polaron-like or dynamically dressed carrier transport, although direct experimental verification of such carrier–lattice complexes remains an important direction for future work.

This behavior differs from the conventional phonon-glass electron-crystal picture in an important way. In the PGEC framework, phonon transport is suppressed while electronic transport is ideally preserved through relatively independent carrier pathways. In $GuaPbI_3$, the present evidence instead points to a softer and more electronically heterogeneous lattice environment, where phonon suppression, electrostatic heterogeneity, and bias-activated transport appear to be closely linked. This does not by itself prove a fully cooperative carrier–lattice transport regime, but it supports the possibility that lattice deformability plays an active role in shaping the observed transport response.

Ultralow lattice thermal conductivity together with electronically active bulk transport in $GuaPbI_3$ can therefore be understood as arising from a lone-pair-rich Pb–I framework coupled to a large polarizable organic cation. Structural softness, electrostatic heterogeneity, and locally deformable lattice environments provide a plausible basis for strong phonon suppression while preserving accessible electronic states near the band edges. More broadly, hybrid frameworks combining lone-pair-active inorganic networks with polarizable organic cations may provide a useful platform for exploring chemically induced lattice softness and ultralow thermal conductivity. From this perspective, materials discovery may benefit from identifying lattice architectures capable of intrinsically suppressing phonon transport without relying solely on nanostructuring or extrinsic disorder engineering.

## 7. Acknowledgement

The authors would like to acknowledge CRF IIT Delhi for providing the required experimental facilities.